\documentclass[jcp,onecolumn,preprint,floatfix]{revtex4-1}
\usepackage[dvipsnames,svgnames,table]{xcolor}
\usepackage{xr}
\usepackage{graphics} 
\usepackage{graphicx} 
\usepackage{color} 
\usepackage{bm} 
\usepackage{bbm}
\usepackage{amsmath}
\usepackage{amssymb}
\usepackage{amsfonts}
\usepackage{rotating}
\usepackage[normalem]{ulem}
\definecolor{DarkYellow}{RGB}{80, 80, 0}

\usepackage[normalem]{ulem}  % for sout command

\begin{document} 

\title{Emergent swimming strategies of a smart three-bead swimmer}

\author{Ruma Maity$^1$}
\email{ruma.maity@tuwien.ac.at}
\author{Maximilian H\"ubl$^{1,2}$}
\author{Julian Lemmel$^3$}
\author{Benedikt Hartl$^{1,4}$}
\author{Gerhard Kahl$^{1}$} 

\affiliation{$^1$Institute for Theoretical Physics, TU Wien, Austria \\$^2${Institute of Science and Technology Austria (ISTA), Am Campus 1, 3400 Klosterneuburg, Austria} \\ $^3$Institute of Computer Engineering, TU Wien \\ $^4$Allen Discovery Center, Tufts University, Medford, MA, USA
%}
}


\begin{abstract}
Low-Reynolds-number microswimmers have recently attracted much interest for their ubiquity in biology and their applications in biotechnology and medicine. However, a key obstacle for the design and deployment of artificial microswimmers lies in their autonomy: to successfully perform tasks in any real-world scenario, these swimmers need to be able to interact with and adapt to their environment without external control. Here, we train a simple two-dimensional model microswimmer (consisting of three-bead) to learn autonomous swimming strategies via Reinforcement Learning, focusing on neuroevolution techniques to derive controller architectures with minimal complexity. We identify five different, characteristic swimming gaits: three of these gaits lead to directed locomotion with varying grades of efficiency and two gaits result in a rotational, inefficient movement. Remarkably, all of these gaits can be achieved by very simple neural networks (with less than ten nodes and weights), showing that low-Reynolds-number swimming can be achieved efficiently and robustly while requiring only minimal ``computational power''. These results are of particular interest to the experimental design of artificial microswimmers and may have implications for modeling biological microorganisms such as \textit{Chlamydomonas reinhardtii}.
\end{abstract}

\date{\today}

\maketitle

%%%%%%%%%%%%%%%%%%%%%%%%%%%%%%%%%%%%%%%%%%%%%%%%%%%%%%%%%%%%%%%%%%%%%%%%%
%%%%%%%%%%%%%%%%%%%%   INTRODUCTION (START)                              

\section{Introduction}
\label{sec:introduction}

Microorganisms belong to the ``low Reynolds number'' world where viscous forces dominate over inertia \cite{Purcell1977}. These microorganisms are abundant in nature and play a key role in many biological phenomena. In an effort to optimize their survival in their diverse habitats these microorganisms have developed a enormous variety of clever strategies which help them to reach a nutrient-rich source \cite{ desai_2017, magar}, to reproduce \cite{ramaswamy}, to escape from a predator \cite{waggett}, to avoid an adverse environment \cite{barclay}, or to hunt prey \cite{stocker} -- to name a few examples. Different microorganisms employ different swimming strategies to propel in their aqueous surrounding: while some microorganisms deform their shape in a non-reciprocal manner \cite{elgeti}, others use different appendages (i.e., cilia, flagella, or archaelle) to swim \cite{dreyfus, pak}: for instance, sperm cells swim with the help of a single flagellum \cite{alvarez, elgeti2}, the bacterium {\it Escheria coli} builds on the rotation of a bundle of appendages \cite{berg}, while the  self-propulsion of the eukaryote {\it paramecium} relies on a wave-like deformation of cilia that are distributed along the body of the cell \cite{hamel, blake1}. 

In an effort to understand the locomotion of microswimmers under low Reynolds number conditions, several models have been proposed \cite{taylor, lighthill, blake, najafi, Johnson, zhao, luo, choi, chen_2021, era_2021, daftari, borra, frydel,rizvi2018}. However, most of these models are endowed by a {\it fixed} swimming strategy \cite{Patra} that inherently lack the essential features necessary to adapt the swimmers' behavior in the incessantly spatially and/or temporarily changing environment -- i.e, they cannot choose an appropriate swimming gait based on their interactions with the dynamically evolving environment. For example, in the case of chemotaxis, a swimmer  needs to adjust its swimming direction accordingly in order to move toward high concentrations of a chemical target to optimize the uptake of nutrition, which -- in real life -- is essential for survival. In general, bio-inspired artificial microswimmers have  promising applications in biomedical fields, such as smart vehicles for targeted drug delivery \cite{Patra, walker,gao, ghosh_mech} or noninvasive surgery \cite{ghosh_mech,nelson}, to name a few examples. However, in  all of these cases, the artificial swimmers might face unpredictable environmental conditions \cite{nassif, mirbagheri} and therefore need to exhibit robust autonomous navigation policies \cite{gao, walker} : the swimmers must be able to adapt to their environment and improvise their swimming strategy in a flexible manner while performing a target task. Rapid  recent developments in the field of Machine Learning have led to various tools and techniques which are  {excellent candidates for the design of such next-generation intelligent microswimmers \cite{cichos_nat, tsang_smart, liu_sr, liu_as, xiong, liu_sm, liu_pof,mohamed,lin_prr, lai, abbasi, nassiri,qin, lu_njp}}. However, approaches to design intelligent controllers for microswimmers have almost exclusively relied on deep Reinforcement Learning (RL) strategies to train complex, large-scale artificial neural networks (ANNs) -- typically with tens or hundreds of thousands of parameters \cite{zou, alonso, salehi}. Although generally showing great promise, RL with deep ANNs can be computationally expensive, operationally brittle, and often do not generalize well to out-of-distribution situations \cite{cobbe2019, henderson2018}, thus limiting its direct applicability to robust, resource-constrained environments that are typical for \textit{in vivo} scenarios.

A previous publication \cite{hartl_2021}  -- dedicated to a simple linear three-bead swimmer model \cite{najafi,Golestanian2008} -- has successfully implemented a genetic algorithm-based RL scheme to \textit{learn} minimally complex strategies for robust and \textit{life-like} microswimmer locomotion and chemotaxis.  In this approach, the microswimmer and its viscous surrounding represented -- in the language of  RL -- the ``environment'', whose movement is stirred by  an ``agent''  with a generally complicated policy that is  substituted by  an ANN.  This \textit{in silico} model relied on a continuous swimming process with discretized agent-environment interactions to autonomously navigate the swimmer. 
In an effort to train the swimmer's policy, a reinforcement scheme is introduced via  a ``reward'' signal $r$ that quantifies task-specific feedback about the quality of the agent's proposed actions in the hydrodynamic environment, such as mean swimming speed or chemotactic efficiency. Through reward maximization via the ``Neuro Evolution of Augmented Topologies'' (NEAT) method \cite{stanley}, surprisingly simple yet robust ANN architectures have been identified, comprising only a handful of sparsely connected neurons. Nonetheless, these evolved ANNs allowed the agent to induce coordinated non-reciprocal body deformations that efficiently navigated the microswimmer in versatile chemotactic environments, generalizing well beyond training conditions and giving rise to emergent \textit{life-like} behavior. Moreover, such simple ANN architectures  make it possible to analyze the internal decision-making strategy of the microswimmer and, due to their simplicity, are promising candidates to design corresponding experimental controllers at the microscale.

In this contribution, we delve into the much more challenging task of two-dimensional swimming of a three-bead triangular swimmer. {At this point it should be noted that meanwhile quite a few contributions have been dedicated to this particular type of swimmers that is able to move in a flexible manner the connecting arms ``around'' the central bead of the swimmer, introducing thereby additional degrees of freedom of bending and angular movement and leading thus to a considerably enhanced configuration space; we refer the reader to a selection of related publication, namely \cite{liu_as, liu_sm, liu_pof}.}
While one-dimensional movement is free from rotation, two-dimensional swimming involves a complex interplay between rotational and translational degrees of freedom. The former opens up a path for diverse swimming gaits suiting different complex environments. Interestingly, the limit cycles of shape deformation may decide the swimming gaits which are completely independent of each others.

Here, we extend the aforementioned one-dimensional model to a two-dimensional three-bead swimmer of triangular shape, with three arms connecting the beads (see Fig.~\ref{fig:introduction}a). This model has originally been introduced to mimic the internal structure of {\it Chlamydomonas} \cite{rizvi2018,Rizvi2018_2,rizvi2019}; the so far most comprehensive study of this model was published by Rizvi {\it et al.} \cite{rizvi2018} identifying a large variety of swimming gaits. The scenario poses significant coordination challenges for the two-dimensional triangular swimmer that is characterized by the arm lengths and the forces acting upon the individual beads. The swimmer can either move in  straight (purely translational) or rotational manner, with a combination of both modes leading to complex swimming behavior, as detailed in this contribution. 
To this end, we use a flexible, transparent reward scheme that involves a weighted linear combination of different partial rewards. Using this scheme in combination with NEAT allows us to emphasize desired target swimming strategies during training -- by tuning the corresponding reward weights -- and to obtain lean solutions for microswimmer controllers in terms of minimally complex ANN architectures. This provides a transparent view of the trained swimming strategies (i.e., the coordination between the changes in the arm lengths and the resulting forces). 

Methodologically, we pair the powerful NEAT algorithm with a novel complex reward scheme for two-dimensional locomotion, which churns out efficient versatile swimming gaits from simplistic ANNs. 
While in traditional RL schemes, the weights of the connections of a fixed ANN are varied,  we  optimize both the structure (topology) and the weights of the networks: 
In a nutshell, NEAT combines a systematic neural architecture search  with corresponding weight updates through evolutionary inspired recombination and mutation operations of topologically related high-quality ANNs (high-reward policies) across generations of candidate solutions; hence the naming ``NeuroEvolution of Augmenting Topologies'' \cite{stanley}. Especially the ability of NEAT to identify minimally complex but highly robust and generalizable agent policies even in rugged reward landscapes renders this method -- for our purposes -- superior to other contemporary works which involve computationally expensive large-scale networks. In general, Evolutionary Algorithms have been refined over several decades \cite{back,eiben}, are presumably among the most prominent candidates to find both versatile and nearly optimal solutions of the most difficult learning problems in computer science and diverse engineering fields. 
This is particularly important in situations where a gradient is difficult to formalize and quantify, say, in a rugged landscape characterized by complex topographical features and irregularities. 
Within NEAT the network topology and the weights of the connections are evolved simultaneously, thus this algorithm has the attractive feature of learning lean functional representations to control dynamic systems, which increases the learning speed, robustness, and generalization capabilities of the networks drastically.

We show that our three-bead swimmer is adaptable and intelligent in the sense that it is able to learn a total of five distinctively different swimming strategies via the RL scheme. They are recovered as the different weights of the contributions to the total reward  are varied at different trial runs; the respective partial rewards consider the coupling of the translational and orientational degrees of freedom as well as geometric aspects. Notably, we could identify three swimming strategies --- referred to as flapping, chiral, and walking modes --- which are characterized by non-reciprocal deformations in the shape of the swimmer \cite{elgeti,najafi_iop, iima, lauga_pre, leshansky}, resulting in a net displacement through space. Each mode involves a specific time-dependent interplay among the three arm lengths (or, alternatively, the forces), and an analysis of these dynamics helps in understanding the respective propagation patterns and decision-making processes of the swimmer. Additionally, two other modes can be identified that are characterized by reciprocal shape deformations. These modes either result in a circular trajectory of the swimmer's center of mass (COM) or induce rotation while maintaining the COM's position fixed. In both scenarios, these swimming strategies do not produce any net movement of the swimmer. 
Our results thus show that combining neuroevolution-based RL with a targeted complex reward scheme allows us to train two-dimensional microswimmers to adapt their swimming strategies in highly viscous hydrodynamic environments, fully compatible with the scallop theorem \cite{Purcell1977}.
In other words, without any prior knowledge of the system, the evolved ANNs provided us the set of forces only from the knowledge of arm lengths which set the swimmer into motion; they represent the decision-making strategies of the swimmer. Therefore, our results are also encouraging for future investigations to be dedicated to more complex microswimmer geometries and to ensembles of microswimmers.

At this point we emphasize that a deeper understanding of the swimming strategies and internal decision-making scenarios of microswimmers is  of high practical relevance: An increasing number of theoretical approaches demonstrate impressive \textit{in silico} navigation strategies in active matter systems. The swimmers are either rigid, with tunable velocities or torques, or deformable and self-propel via coordinated body transformations \cite{takatori2025, volpe2024, mo2023, cichos_nat}.  
Moreover, recent trends in developmental robotics and collective intelligence have started to utilize decentralized design principles to optimize both the form and function of bio-inspired robotic agents \cite{mcmillen, pontes2022, sudhakaran2021, pathak2019, brodbeck2015, bongard2011, lungarella2003}.
Again, through neuroevolution techniques, extremely lean controllers of composite microswimmers have been developed that are highly robust against environmental perturbations, generalize across body shapes, and can handle severe morphological damage (partially dysfunctional or missing components) \cite{hartl_nature}. 
Moreover, such neuroevolution-based \textit{in silico} design principles have given rise to experimental realizations of bio-hybrid ``living robots'', termed Xenbotos \cite{kriegman2020, kriegman2021} and Antrhobots \cite{gumuskaya2024}, which show great therapeutic potential, e.g., in neural wound healing. 
In this context, bio-inspired machine learning tools (as the ones used in this contribution, namely, NEAT) are excellent candidates to train  artificial microswimmers to achieve long--term goals, such as targeted drug delivery \cite{park, bunea}, chemotactic navigation \cite{maity_epje, maity_pof, maity_jfm, jin_pnas}, and, in general, artificial life \cite{langton}.

The manuscript is organized as follows: in Section II we present our three-bead swimmer model and the underlying equations-of-motion in the low-Reynolds number regime; further, we briefly present the NEAT scheme that we use to train the swimmer and discuss the implementation of the model within the RL scheme. Section III is dedicated to the presentation and the discussion of our results. The manuscript is closed by concluding remarks. A Supplementary Information (SI) is provided which contains additional information about our investigations.

%%%%%%%%%%%%%%%%%%%%   INTRODUCTION (END)                                 
%%%%%%%%%%%%%%%%%%%%%%%%%%%%%%%%%%%%%%%%%%%%%%%%%%%%%%%%%%%%%%%%%%%%%%%%%

%%%%%%%%%%%%%%%%%%%%%%%%%%%%%%%%%%%%%%%%%%%%%%%%%%%%%%%%%%%%%%%%%%%%%%%%%
%%%%%%%%%%%%%%%%%%%%   MODEL AND METHODS (START)                        

\section{Model and methods}
\label{sec:models_methods}

\subsection{Microswimmer model and dynamics} 
\label{subsec:models}

\begin{figure}
    \centering
    \includegraphics[width=\textwidth]{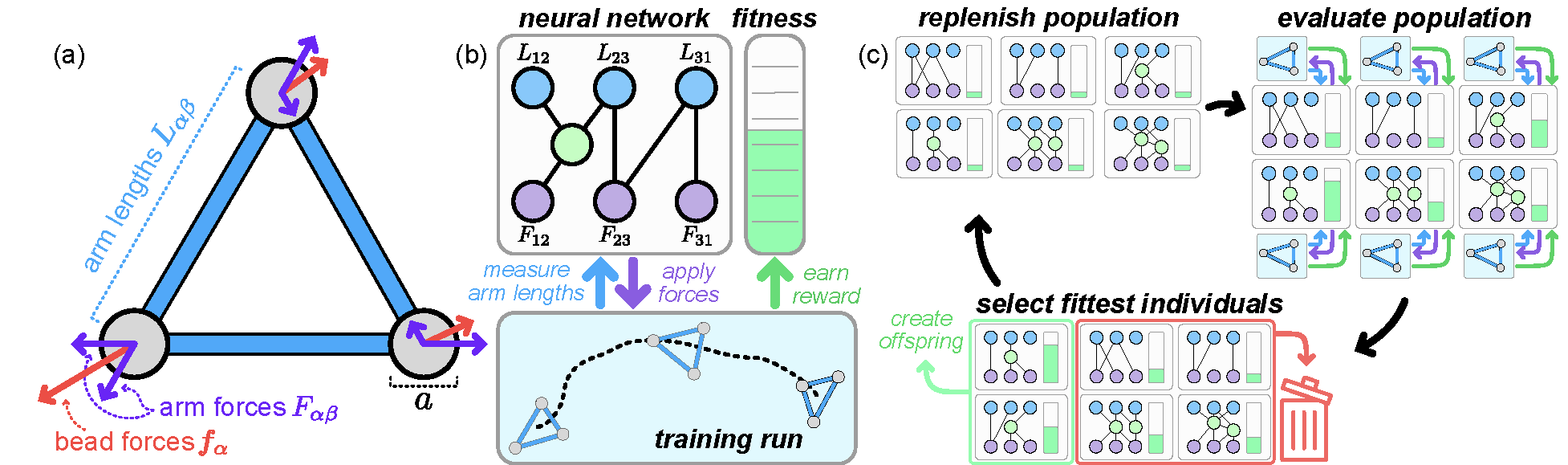}
    \caption{Microswimmer model and training setup. (a) Sketch of the microswimmer model, consisting of three-bead of radius $a$, connected by three arms. The arms can exert forces on the beads that, coupled with the hydrodynamic interactions between the beads, lead to locomotion of the swimmer. (b) Interplay between the neural network and the swimmer simulation. Over the course of a training run, the neural network repeatedly measures the swimmer arm lengths, from which it computes arm forces that are then applied to the swimmer's arms. Once the training run is completed, the network's fitness is updated based on the amount of reward it collected. (c) Schematic illustration of the optimization loop. A given collection of networks is evaluated by performing the procedure outlined in (b). Only the fittest individuals are retained; from these, a new population of networks is created via random mutation and recombination.}
    \label{fig:introduction}
\end{figure}

In this paper we investigate a simple microswimmer model originally introduced in~\cite{rizvi2018}, consisting of three-bead immersed in a viscous fluid and connected by three actuators, or ``arms'', as shown in Fig.~\ref{fig:introduction}(a). The arms can apply forces to the beads via expansions and contractions; however since the forces an arm can apply on its two attached beads are always equal and opposite, the total net force and the total net torque on the swimmer remain zero, i.e., $\sum_\alpha\mathbf{f}_\alpha = 0$ and $\sum_\alpha\bm{\Lambda}_\alpha = 0$. Here, $\mathbf{f}_\alpha$ and $\bm{\Lambda}_\alpha$  are respectively the total force and torque on the bead $\alpha$ with henceforth $\alpha = 1,2,3$. Still, the beads do interact with each other through the surrounding fluid and this hydrodynamic coupling makes it possible for the swimmer to move even in the absence of a net force -- as long as the arms perform suitable ``swimming strokes''.

Denoting the positions of the beads by $\bm{x}_\alpha$ 
and their (uniform) diameters 
by $a$, we have the following equation of motion:

\begin{equation}
\label{eom}
    \frac{d \bm{x}_\alpha}{dt} = \mu \bm{f}_\alpha (\mathbf{x}_\alpha, t) + \sum_{\beta \neq \alpha} \mathbf{G}_{\alpha\beta} \bm{f}_\beta (\mathbf{x}_\beta, t) \,,
\end{equation}
where, $t$ is time, $\mu = (3\pi \eta a)^{-1}$ is the bead mobility, $\eta$ is the viscosity of the fluid, and,  $\mathbf{G}_{\alpha\beta} = \mathbf{G}(\bm{x}_\beta - \bm{x}_\alpha)$ is the Oseen tensor, which causes the hydrodynamic coupling between the beads and is given by
\begin{equation}
\label{oseen}
    \mathbf{G}_{\alpha \beta} = \frac{1}{8\pi \eta} \Bigg(\frac{\mathbf{I}}{x_{\alpha \beta}} + \frac{\mathbf{x_{\alpha \beta}} \otimes \mathbf{x_{\alpha \beta}}}{x_{\alpha \beta}^3}\Bigg)\, ,
\end{equation}
$\mathbf{I} $ is the identity matrix. $\otimes$ denotes the dyadic product. Denoting the tensile force on the arm connecting bead $\alpha$ and $\beta$ by $F_{\alpha\beta}$ ($F_{\alpha\beta} > 0$ corresponds to expansion, $F_{\alpha\beta} < 0$ to contraction), we can write the force on bead $\alpha$ as

\begin{align}
    \bm{f}_\alpha &= \sum_{\beta \in \mathrm{neighs}(\alpha)}[F_{\alpha\beta} + D_{\rm R}(|\bm{x}_\alpha - \bm{x}_\beta|)]\frac{\bm{x}_\alpha - \bm{x}_\beta}{|\bm{x}_\alpha - \bm{x}_\beta|} .
\end{align}
where the sum runs over all beads neighboring bead $\alpha$, and $D_\mathrm{R}(x)$ is a restoring force that prevents very large deformations (see Supplementary Material -- SI) for more details. We use the notation  $|\mathbf{x}_\alpha - \mathbf{x}_\beta| = |\mathbf{x}_{\alpha\beta}| = L_{\alpha\beta}$.

\subsection{Learning to swim} 
\label{subsec:learning}

For the swimmer to swim, the arm forces $F_{\alpha\beta}$ have to be prescribed in a suitable manner. This can, in principle, be achieved simply by choosing the forces to be appropriate, periodic functions of time, $F_{\alpha\beta} \equiv F_{\alpha\beta}(t)$~\cite{rizvi2018,Rizvi2018_2,rizvi2019,rizvi_prf,ziegler_jop}. However, it is often unclear how to find the appropriate forces $F_{\alpha\beta}(t)$ to achieve the desired swimming behavior; and it is even harder to identify optimal swimming strategies~\cite{rizvi2018, ziegler_jop}. Even more importantly, an external, fixed prescription cannot represent a robust strategy for locomotion in any real-world setting, since it does not take the constantly changing environment or even the current state of the swimmer into account. 

A more realistic and robust way to achieve locomotion is to make the forces a function of the bead positions, or equivalently of the current arm lengths, $F_{\alpha\beta} =F_{\alpha\beta}(L_{12}, L_{23}, L_{31})$.
%, where $L_{\alpha\beta} = |\bm{x}_\beta - \bm{x}_\alpha|$.
This way, Eq.~\eqref{eom} becomes a closed equation in $\bm{x}_\alpha$, and periodic swimming strokes can emerge as limit cycles in the solution space (see on a later stage Fig. \ref{fig:triag_comb_length} and the related discussion). Thus, if we engineer $F_{\alpha\beta}(L_{12}, L_{23}, L_{31})$ appropriately, we should be able to make the swimming behavior and the swimming strategies of the swimmer much more robust and responsive to external perturbations, or even allow the swimmer to change behavior in an autonomous fashion based on external cues. 

However, we are still dealing with the issue of how to choose the functions $F_{\alpha\beta}(L_{12}, L_{23}, L_{31})$ in an appropriate manner such that the swimmer can decide how to propagate in an autonomous and efficient manner. Here, we solve this issue using function approximators from machine learning, i.e.,  artificial neural networks (ANNs), to parameterize the possible functions (i.e., from the bead positions to the arm forces). We optimize these neural networks using Reinforcement Learning (RL) approach based on a genetic algorithm, specifically the NeuroEvolution of Augmenting Topologies (NEAT) approach \cite{stanley, neat-python}. Importantly, we put a large emphasis on simplicity and efficiency in the size and the internal architecture of the ANNs.  As shown below, our resulting networks are tiny and extremely simple, most containing less than ten internal weights and hidden nodes. {For a short summary of the NEAT algorithm we refer the reader to Section I.A of the Supplementary Information; as for the numerical parameters that we have used in applying this method we refer the reader to the Supplementary Information of Ref. \cite{hartl_2021}.}

The basic structure of our approach is sketched in Fig.~\ref{fig:introduction}(b-c). As stated above we use the NEAT algorithm~\cite{neat-python, stanley} to initialize a population of 200 neural networks. Each network takes as input the current lengths of the swimmer's arms and returns the (approximated) forces on the arms. The networks initially contain no hidden nodes and randomly initialized weights, meaning that they effectively return random results to begin with. We limit the forces output by the networks to be in the range $(-F_0, F_0)$.

We then evaluate each network by integrating the equations of motion -- Eq.~\eqref{eom} --, letting the network return arm forces, and measuring how well the network has achieved its objective, namely to enable the swimmer to swim successfully and propel in an autonomous fashion in the medium. To this end, we initialize the swimmer as an equilateral triangle with small, random spatial perturbations. At every interaction time step of $t_\mathrm{int} = \eta a^2 / F_0$\footnote{$t_{\rm int} = T_0$ is the interaction time scale that provides a rough estimate of the stroke period of a swimmer. $t_{\rm int}$ is furthermore the time step of our calculations, setting thus a limit on how fast the arms of the swimmer can be expanded or contracted.}, we feed the current arm lengths into the network, and then apply the returned arm forces from the network output until the next interaction step. We run all training simulations for 6000 interaction steps, and then assign a \emph{fitness} value based on the prescribed reward scheme to each network that quantifies how well it achieved our desired objective as stated above (see Fig.~\ref{fig:introduction}(b)).

Once all networks in the population have been evaluated the networks are ranked by their fitness; we consider a term a collection of ANNs which represents a single iteration running over $6000$ interaction steps of the respective iteration as ``generation''. Low-performing networks (which can not return arm forces that render the swimmer motile in the desired manner) are discarded; on the other hand, the highest-performing networks are randomly mutated and recombined to form new networks (via a so-called ``crossover'' operation). The initializations of the these run is randomised which, in turn,  gives rise to randomised results. Taking an average over 30 runs cancels out possible anomalies and provides us the underlying true behaviour. The new ANNs created in this manner form the next population, which is again evaluated, as sketched in Fig.~\ref{fig:introduction}(c). This evolutionary optimization is carried out over 1000 ``generations''. In the end, the network with the highest fitness defines the ``optimal'' swimming strategy, which is associated to the swimmer with the most efficient motility subjected to the respective reward scheme. For more technical details we refer the reader to the SI.

\subsection{Teaching to swim}
\label{subsec:teaching}

We control the outcome of this training by defining how the fitness value is assigned which encodes our desired objective. In the language of RL, each network collects a \emph{reward} during evaluation and the fitness value is the sum of the total collected reward.
Since it is our goal to uncover different swimming gaits (including both translation and rotation) and the underlying learning and coordination strategies, we introduce a flexible reward that is a weighted sum of different contributions, each weighted by a parameter. Varying in a systematic manner these weights and searching for the minimal solutions of the ANNs (in terms of their internal architecture) we can unambiguously trace back the impact of the partial weights on the particular strategies that lead, in turn, to well defined swimming gaits. That way, we can systematically explore the space of swimming gaits by varying the weights of the different contributions.

Our reward signal has the form 

\begin{equation}
    r = \rho_1 r_1 + \rho_2 r_2+ \rho_3 r_3 + \rho_4 r_4 \,,
    \label{reward_total}
\end{equation}
with the partial rewards $r_i$

\begin{align*}
    r_1 &= \tau \, |\mathbf{v}_\mathrm{COM}(T)|, &&r_3 = \min_{\alpha}(|\bm{x}_\alpha(T) - \bm{x}_\mathrm{COM}(T)|) \,,\\
    r_2 &= \tau \,\frac{1}{T}\int_0^T |\bm{v}_\mathrm{COM}(t)|\,dt\,, &&r_4 = \frac{\eta a^3}{F_0} \sum_\alpha \frac{|(\mathbf{x}_\alpha(T) - \bm{x}_\mathrm{COM}(T)) \times \mathbf{v}_\alpha(T)|}{\ell^2} \,;
\label{partial_rewards}    
\end{align*}
in the above relations $\rho_i$ are the respective weights of the partial rewards $r_i$ with constraint,
$$
\sum_{i=1}^4 \rho_i = 1\, .
$$
$\bm{v}_\mathrm{com}(t)$ is the center of mass (COM) velocity of the swimmer at time $t$, $\bm{x}_\mathrm{com}$ is the COM position, $\bm{v}_\alpha(t)$ is the velocity of bead $\alpha$, and $\ell$ is the average value over the three arm lengths of the swimmer at $t=0$.  $\tau$ is the time interval for which the predicted displacement (via the partial reward $r_1$) is calculated, based on the instantaneous velocity of the COM of the swimmer $(\mathbf{v}_\mathbf{com}(T))$ at the final time $T$; the partial reward $r_2$ is based on to average velocity of the COM of the swimmer have been calculated. The partial reward $r_3$ picks out the smallest distance of the beads with respect to the COM of the swimmer and $r_4$ couples the translational and the rotational degrees of freedom of the swimmer.

Each reward term favors different behaviors: $r_1$ supports the swimmer's tendency for a continuous movement with a higher velocity, but does not distinguish between curved or straight-line motion. $r_2$ rewards a preferably long distance, measured from the starting point; thus it favors locomotion via a translational (but not necessarily linear) motion. $r_3$ is a geometric reward which penalizes large deformations and in particular prevents the swimmer from collapsing into a line and moving in one dimension. $r_4$ is related to the angular velocity of the beads around the center of mass, and favors rotational motion, as it occurs in different swimming gaits discussed in Sec. \ref{sec:results}.

Without going into too many details, we exemplify the flexibility of the above scheme for the total reward with a few examples: (i) for dominating values of $\rho_1$ and $\rho_4$ we observe a pure rotational motion of the swimmer; (ii) on the other hand, if $\rho_2 \gg \rho_1, \rho_2, \rho_3$, this choice of weights ensures a translational propagation and three uniquely identified swimming gaits (see Sec. \ref{sec:results}); (iii) in absence of $r_3$ (i.e., for $\rho_3 = 0$) the swimmer risks to transform into a linear, one-dimensional three-bead swimmer. 

In the following Section we now explore the swimming strategies that emerge if we tune the weights of the different reward terms.

%%%%%%%%%%%%%%%%%%%%   MODEL AND METHODS (END)                        
%%%%%%%%%%%%%%%%%%%%%%%%%%%%%%%%%%%%%%%%%%%%%%%%%%%%%%%%%%%%%%%%%%%%%%%%%

%%%%%%%%%%%%%%%%%%%%%%%%%%%%%%%%%%%%%%%%%%%%%%%%%%%%%%%%%%%%%%%%%%%%%%%%%
%%%%%%%%%%%%%%%%%%%%   RESULTS (START)                        

\section{Results}
\label{sec:results}

In our investigations we have systematically varied the coefficients $\rho_i$ of the total reward -- see Eq. (\ref{reward_total}). Since, the proposed reward scheme is complex in nature, the convergence of the training is rather noisy. The network configurations have a $\tanh(x)$ activation function. By carefully analysing the trajectories generated by the emerging microswimmers (for details see SI) we could identify the following five swimming strategies (or swimming modes):  (i) flapping mode, (ii) chiral mode, (iii) walking mode, (iv) rotational mode, and (v) circular mode. 

The relevant quantities that characterize the swimming gaits of the microswimmer
that will be discussed and displayed in the following are -- all of them as functions of time $t$ -- the arm lengths $L_{\alpha \beta} = L_{\alpha \beta} (t)$ and the forces $F_{\alpha \beta} = F_{\alpha \beta} (t)$ acting on the arms that connect beads $\alpha$ and $\beta$. Lengths, forces and time are given in units of the diameter of the beads, $a$, $F_0$, and $T_0 = \eta a^2/F_0$, respectively; the viscosity $\eta$ is set to unity. 

As discussed below the first three  swimming modes are characterized by {\it non-reciprocal} deformations of the shape of the swimmer leading thus to a net displacement in space of the COM of the swimmer. In contrast, the rotational and the circular modes do not share this feature of shape deformations of the swimmer. Thus -- and in full agreement with the scallop theorem \cite{Purcell1977} -- the swimmer does not propagate in these swimming modes in space but performs, instead, an inefficient motion with a strongly localized trajectory of the (COM) of the swimmer. 

\begin{figure}
\centering
%%%%%%%%%%%%%%%%%% please insert correct name
\includegraphics[width=0.82\linewidth]{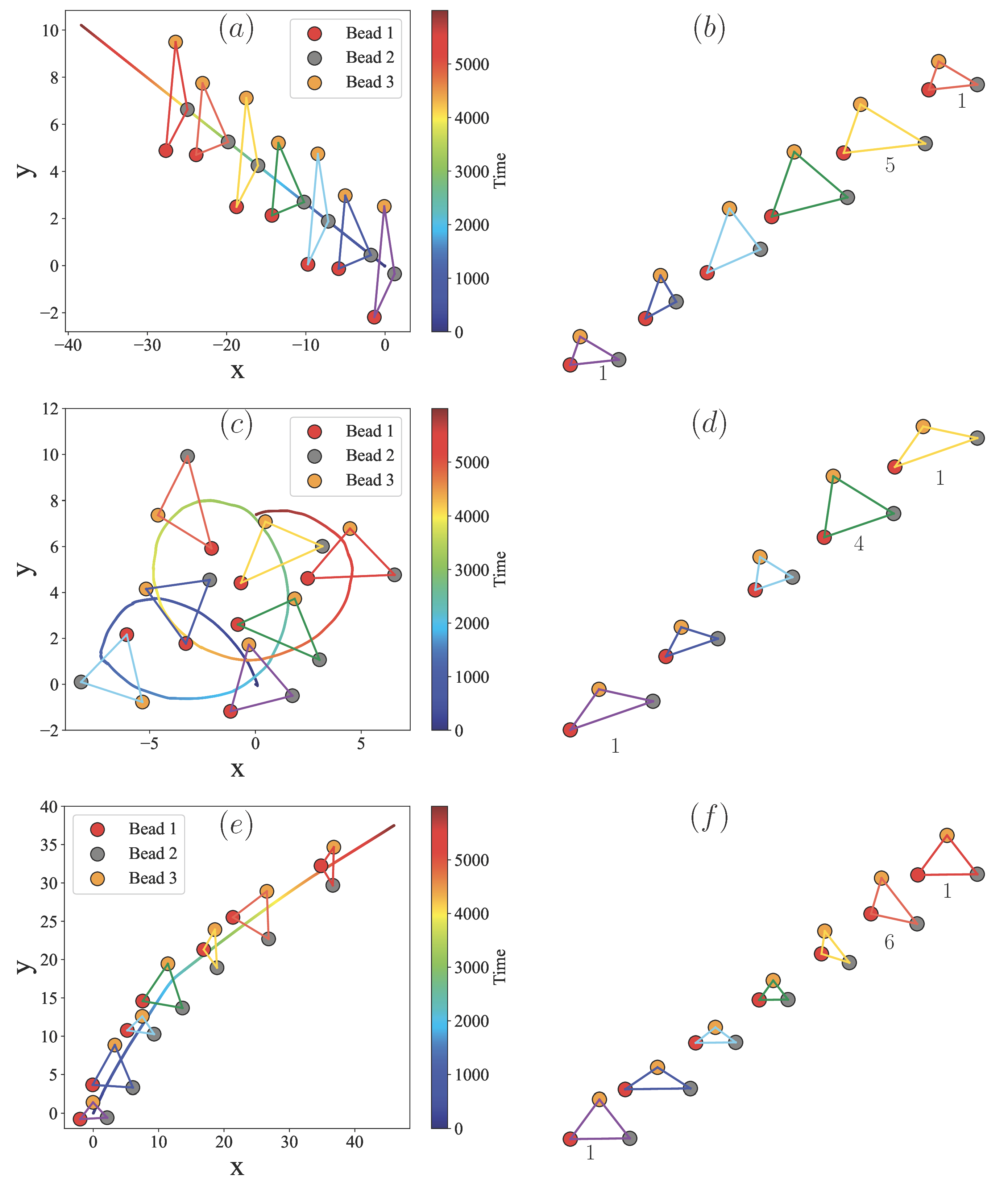}
%%%%%%%%%%%%%%%%%%
\caption{Visualization of the characteristic features of the flapping (panels (a) and (b)), the chiral (panels (c) and (d)), and the walking (panels (e) and (f)) swimming gaits of the triangular swimmer. Panels in the left column (i.e., panels (a), (c), and (e)): trajectories of the triangular swimmer in the $(x, y)$-plane as a function of time (from ``blue'' to ``red'' -- see colour coded time scale next to the panels). For clarity each bead of the swimmer is characterized by a specific colour (as labeled). 
Panels in the right column (i.e., panels (b), (d), and (f)): visualization of the non-reciprocal arm deformations for the three different swimming gaits (specified above) over a period of motion; the numbers label the strokes of the periodically repeated deformation of the swimmer (see also text). The size of the beads is not scaled with the arm lengths.}
\label{fig:triag_trajectories}
\end{figure}

The relevant data of the first three swimming strategies are summarized in Figs. \ref{fig:triag_trajectories} to \ref{fig:triag_comb_length}: Fig. \ref{fig:triag_trajectories} displays the trajectory of the swimmer (displaying both beads and connecting arms) over an extended time interval of 6000 time steps (left panels) and the strokes that characterize each swimming pattern (right panels) which are repeated periodically. Fig. \ref{fig:triag_f_l_2d} shows the forces $F_{\alpha \beta}$ (left panels) and the arm lengths $L_{\alpha \beta}$ (right panels) as functions of time in a two-dimensional plot while Fig. \ref{fig:triag_comb_length} displays these quantities in a three-dimensional representation along with the projections of the three-dimensional curve on the respective two-dimensional subspaces. In Fig. \ref{fig:triag_long_traj} we display the trajectories due to the {\it trained} networks over a longer time period for these three swimming strategies. As a supplementary information we have prepared movies for each of the three aforementioned modes which show the development of trajectories of the microswimmer over 4000 time steps. 

Related figures for the two other (inefficient) swimming strategies are presented in Figs. \ref{fig:triag_trajectories_A} -- typical trajectories and strokes and Fig.  \ref{fig:triag_f_l_2d_A} -- $F_{\alpha \beta}$ and $L_{\alpha \beta}$ as functions of time. 

The weights $\rho_i$ of the partial rewards that contribute to the total reward via Eq. (\ref{reward_total}) are summarized in the Table.

\begin{table}[t!]
\begin{tabular}{|l|c|c|c|c|} \hline
mode        & $\rho_1$ & $\rho_2$ & $\rho_3$ & $\rho_4$ \\ \hline
flapping    & 0.025    & 0.750    & 0.100    & 0.125    \\ \hline
chiral      & 0.050    & 0.750    & 0.175    & 0.025    \\ \hline
walking     & 0.075    & 0.750    & 0.025    & 0.150    \\ \hline
rotational  & 0.100    & 0.750    & 0.050    & 0.100    \\ \hline
circular    & 0.075    & 0.750    & 0.100    & 0.075    \\ \hline
\end{tabular}
\caption{Weights $\rho_i$ of the partial rewards $r_i$, that add up -- via Eq. (\ref{reward_total}) -- to the total reward $r$ for the swimming modes listed in the first column. 
}
\label{tab:coefficients}
\end{table}

\subsection{Flapping mode}
\label{subsec:flapping_mode}

The most efficient swimming strategy in the sense of a long trajectory is the flapping mode (which occurs in the majority of training events) which is entirely free of rotational motion. The typical trajectory of a swimmer in this mode is depicted in panel (a) of Fig. \ref{fig:triag_trajectories}; it consists of five strokes which are periodically repeated -- see panel (b) of this Figure with the respective numbering of the strokes. The swimmer maintains during this mode the geometry of a isosceles triangle whose shape changes from stroke to stroke. All these features are depicted in panel (b) of Fig. \ref{fig:triag_f_l_2d}: the edges of the blue coloured pentagon reflect the five strokes, the linear projection of this curve onto the $(L_{12}, L_{23}$)-plane verifies that the swimmer maintains the shape of an isosceles triangle. The latter feature is also nicely depicted in panel (a) of Fig. \ref{fig:triag_f_l_2d} where the curves $L_{12}$ and $L_{23}$ are found to be identical: these two arm lengths vary as functions of time in an independent manner of $L_{13}$; this arm length oscillates in a periodic pattern that is shifted from the other $L$'s by one stroke, establishing thereby the non-reciprocal feature of this swimming gait. The forces $F_{\alpha \beta}$ show in their three-dimensional representation a remarkably simple picture of a rectangular-shaped cycle with one of the corners being degenerate; this corner can easily be identified from panel (a) of Fig. \ref{fig:triag_f_l_2d} as the position where all $F_{\alpha \beta}$ coincide in one single point. This swimming strategy which is free of any rotation allows the swimmer to move forward in a highly efficient manner, i.e., the COM propagates indeed along a straight line, being thus the most efficient swimming strategy observed: it does indeed cover thus the longest distances as compared to the other swimming strategies (see also Fig. \ref{fig:triag_long_traj} where the trajectory of the swimmer is shown over 30~000 time steps); note that the COM performs a tiny oscillatory motion along the otherwise straight trajectory. 

The related ANN is depicted in panel (a) of Fig. 2 of the SI: it is surprisingly simple as it does not contain any internal neurons and only five straight connections between the input and the output layers. 

\subsection{Chiral mode}
\label{subsec:chiral_mode}

\begin{figure}
\centering
%%%%%%%%%%%%%%%%%%%%% please insert correct name
%\includegraphics[scale = 1.4]{figs/c_len_force_sche_apr_25.pdf}
%%%%%%%%%%%%%%%%%%%%%%%

\includegraphics[scale = 1.4]{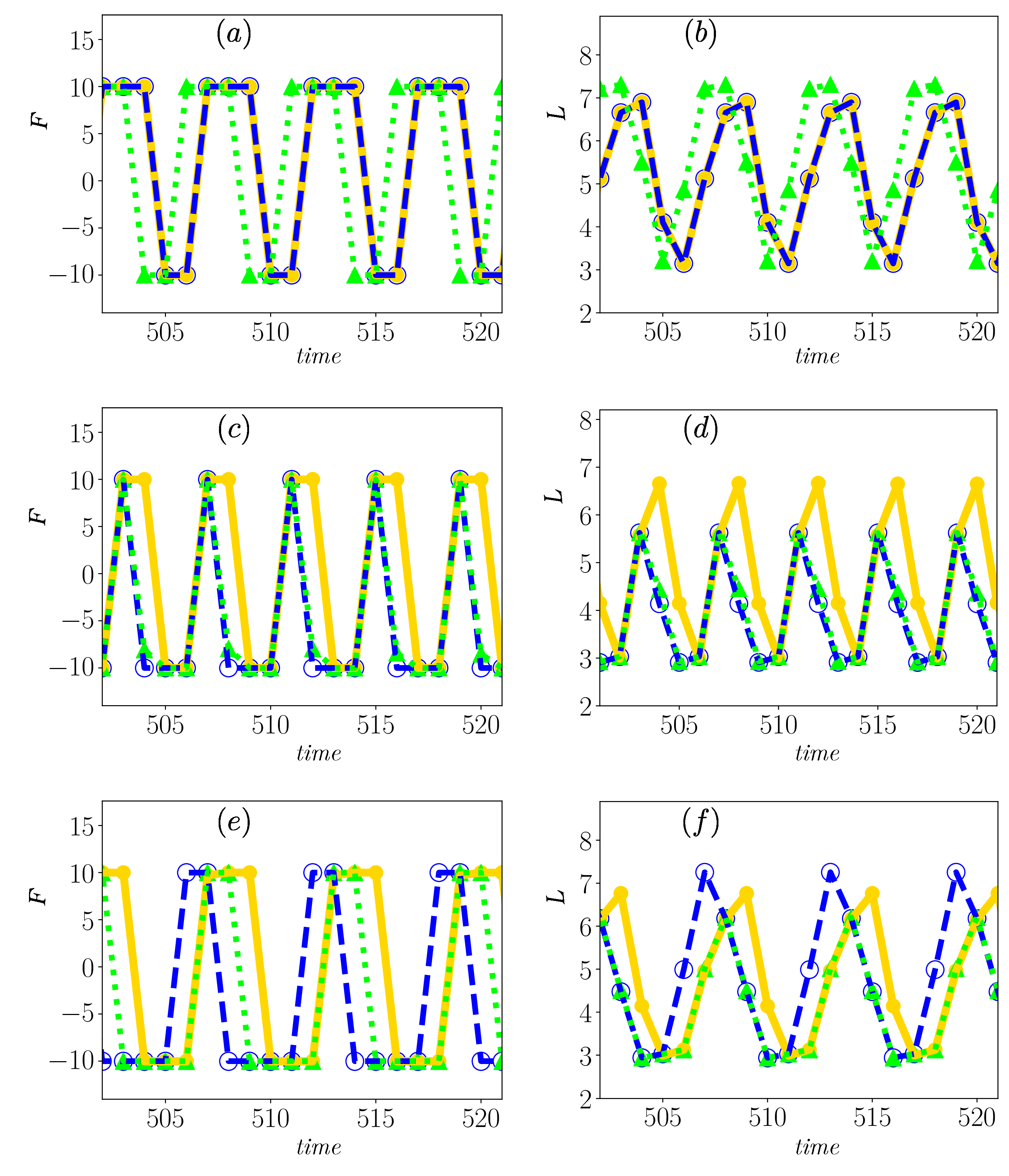}
%%%%%%%%%%%%%%%%%%%%%%%
\caption{Forces $F_{\alpha\beta}$ (panels (a), (c), and (e); as labeled) and arm lengths $L_{\alpha\beta}$ (panels (b), (d), and (f); as labeled) as functions of time over an arbitrarily chosen time window of stationary propagation of the triangular swimmer. Panels in the top row: flapping mode, panels in the central row: chiral mode, and panels in the bottom row: walking mode. Here, $F_{12}$, $F_{23}$, and $F_{13}$ are indicated by,
%\begin{tikzpicture}[baseline=-0.5ex]
 % \draw[Goldenrod, solid, line width=3pt] (0,0) -- (1cm,0);
%\end{tikzpicture} 
(Yellow, solid),
%\begin{tikzpicture}[baseline=-0.5ex]
 % \draw[blue, dashed, line width=1.9pt] (0,0) -- (1cm,0);
%\end{tikzpicture} 
(blue, dashed), and
%\begin{tikzpicture}[baseline=-0.5ex]
 % \draw[green, dotted, line width=1.5pt] (0,0) -- (1cm,0);
%\end{tikzpicture} 
(green, dotted) 
lines, respectively.
\label{fig:triag_f_l_2d}}
\end{figure}

\begin{figure}
\centering
%%%%%%%%%%%%%%%%%%%% please insert correct name
\includegraphics[scale = 2.2]{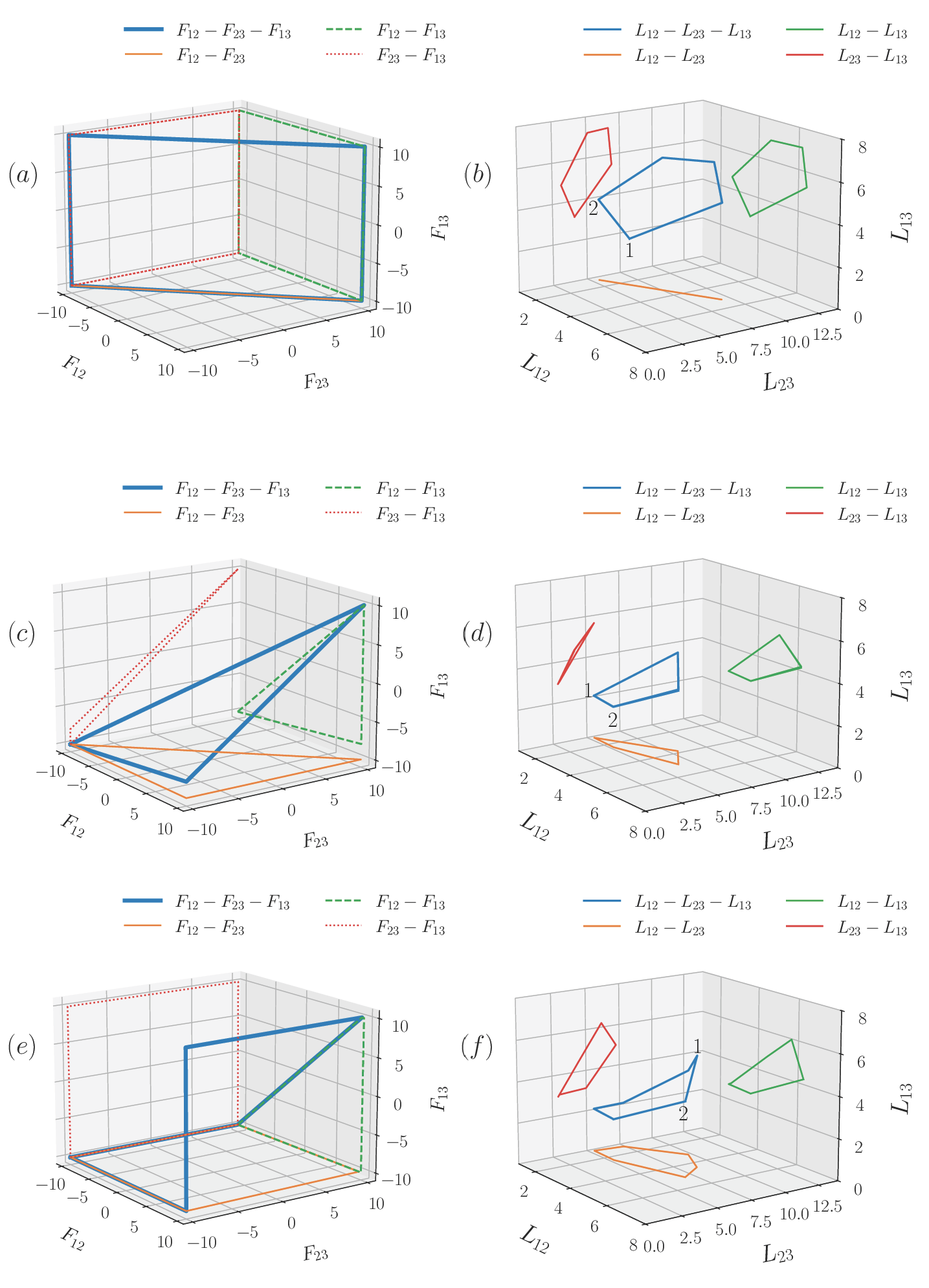}
%%%%%%%%%%%%%%%%%%%%%
\caption{Forces $F_{\alpha\beta}$ (panels (a), (c), and (e); as labeled) and arm lengths $L_{\alpha\beta}$ (panels (b), (d), and (f); as labeled) in their respective three-dimensional representations (blue curves). The red, green, and orange curves represent projections of the blue curve onto the related two-dimensional subspaces (as labeled). Curves are shown over a periodically repeated stroke pattern, displayed for an arbitrarily chosen time window of stationary propagation of the triangular swimmer. Panels in the top row: flapping mode, panels in the central row: chiral mode, and panels in the bottom row: walking mode.}
\label{fig:triag_comb_length}
\end{figure}

The chiral mode is again remarkably simple and consists of four strokes which are repeated periodically; for the visualiation of the trajectories and of the strokes see panels (c) and (d) of Fig. \ref{fig:triag_trajectories}, respectively. In this mode the triangular swimmer maintains essentially again an iscosceles shape: this is confirmed by the identical time-dependencies of $L_{23}$ and $L_{13}$, observable in panel (d) of Fig. \ref{fig:triag_f_l_2d} and the essentially linear projection of the three-dimensional representation of the strokes in panel (d) of Fig. \ref{fig:triag_f_l_2d}. At the beginning of one circle -- say where the $L_{\alpha \beta}$ attain their minimum values -- the three arm lengths expand via identical curves along the first stroke -- see panel (d) of Fig. \ref{fig:triag_f_l_2d}. While $L_{23}$ and $L_{13}$ shrink along the subsequent second stroke (again along identical curves) the arm length $L_{12}$ continues to grow and only starts to shrink as the third stroke sets in; eventually all three arm length assume the same value at the end of the last stroke and the subsequent circle sets in.  This complex interplay of the time-dependencies of the $L_{\alpha \beta}$ eventually lead to a rotational motion which is superposed to the linear propagation. Similar as for the arm lengths also the three curves of the related forces coincide as functions of time along the first stroke; while forces $F_{23}$ and $F_{13}$ diminish during the second stroke, $F_{12}$ remains constant along the second stroke and only starts to diminish as the third stroke sets in. All these features are also nicely reflected in the three-dimensional presentations of the $L_{\alpha \beta}$ and the $F_{\alpha \beta}$ -- see Fig. \ref{fig:triag_f_l_2d};  note that due to the aforementioned features of the $F_{\alpha \beta}$ as functions of time one corner of the three-dimensional representation of the forces degenerates (reducing thus to a triangular shape) -- see Fig. \ref{fig:triag_f_l_2d}. On a larger scale (i.e., over an observation window of 30~000 time steps -- see Fig. \ref{fig:triag_long_traj} -- we get an impression of the complex propagation mode within this swimming strategy.

The related ANN shows again a very simple internal architecture which is free of internal nodes and contains six straight, internal connections between the input and the output layers.

\subsection{Walking mode}
\label{subsec:walking_mode}

The walking mode is undoubtedly the most complicated swimming gait among the non-reciprocal modes that we have identified. A representative section of the trajectory of this mode is depicted -- along with the related six strokes -- in panels (e) and (f) of Fig. \ref{fig:triag_trajectories}. The walking mode is the consequence of a complex interplay of the arm lengths $L_{\alpha \beta}$ (and of the forces $F_{\alpha \beta}$) as functions of time. Starting the sequence of strokes -- for instance -- at the minimum of $L_{23}$ (blue curve) we can see that -- while $L_{23}$ expands -- that $L_{12}$ and $L_{13}$ maintain essentially their short lengths. In the next stroke these two arms start to expand concomitantly during strokes two and three. At the end of the latter one, $L_{23}$ has  meanwhile reached its maximum and now experiences a contraction; this arm continues to decrease, now synchronized with $L_{13}$. Meanwhile $L_{12}$ continues to expand along one more stroke, but then drops during two subsequent strokes, regaining the value of $L_{13}$ and defining thus the end of one cycle of the walking mode. The forces $F_{\alpha \beta}$ show the following rather intricate time-dependence: $F_{13}$ remains first constant at its minimum value during four strokes; then this force grows rapidly within one stroke to its maximum value which it maintains during two strokes. In contrast, $F_{23}$ shows a reversed time-dependence, i.e., it remains for two strokes at its minimum value and then four strokes at the maximum value. Eventually, $F_{12}$ shows a regular sequence, maintaining during three strokes its maximum and three strokes at the minimum value, respectively. This complex time-dependencies induce that the forces $F_{12}$ and $F_{23}$ increase concomitantly during one stroke. The related three- and two-dimensional representations of the $L_{\alpha \beta}$ and $F_{\alpha \beta}$ show related complex shapes of polygons -- see panels (e) and (f) of Fig. \ref{fig:triag_f_l_2d} . On a larger scale (see Fig. \ref{fig:triag_long_traj} we observe that the swimmer propagates in the chiral mode for a while along a straight line, then some small reorientation sets in and the swimmer continues along another straight line, and so forth.

In contrast to the other previously discussed swimming modes the emerging network of the walking mode shows a rather complex internal architecture: the ANN has three internal nodes and twelve internal connections -- see panel (e) of Fig. 2 of the SI.

\subsection{Rotational mode}
\label{subsec:rotational}

In the rotational mode the beads orbit around the position of the COM of the swimmer (see panel (a) of Fig. \ref{fig:triag_trajectories_A}) which is fixed in space. The mode consists of three strokes which are repeated periodically. Along such a rotational cycle one arm length, say $L_{12}$ (see panel (b) of Fig. \ref{fig:triag_f_l_2d_A}), is kept at a fixed length; meanwhile the other two arm lengths perform a periodic sequence of contractions and expansions as functions of time, with the related oscillations being shifted by one stroke. The resulting swimming strategy thus does not lead to a net propagation -- in full agreement with the scallop theorem \cite{Purcell1977}. An identical time-dependence is observed for the forces $F_{\alpha \beta}$ as displayed in panel (a) of Fig. \ref{fig:triag_f_l_2d_A}. In the three-dimensional representation of the $F_{\alpha \beta}$ and the $L_{\alpha \beta}$ we find very simple isosceles triangular shapes, the number of three corners being imposed by the number of strokes; this observation also holds for the related projections on the two-dimensional subspaces.

\subsection{Circular mode}
\label{subsec:circular}

\begin{figure}
\centering
\includegraphics[scale = 1.4]{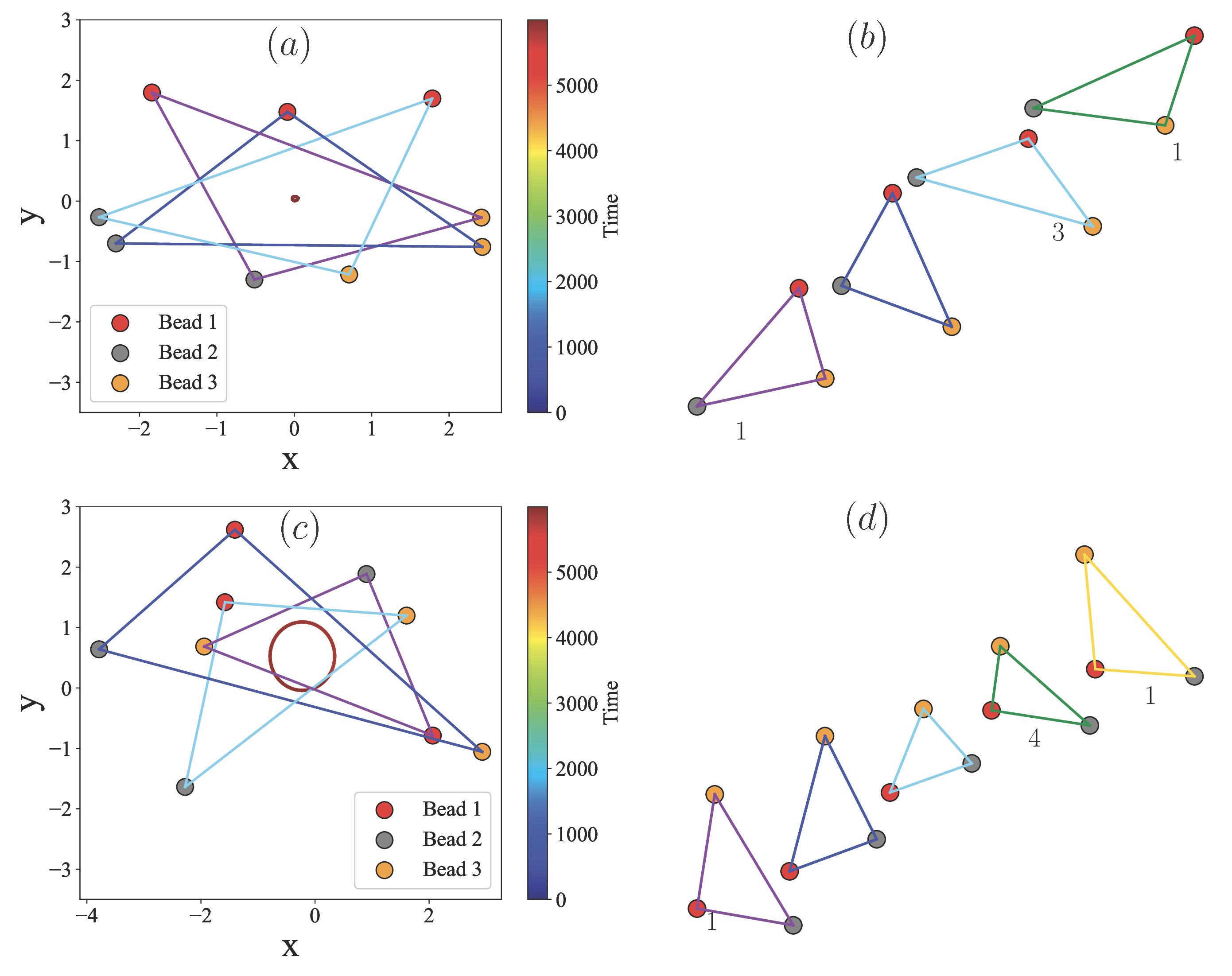}
\caption{Visualization of the characteristic features of the rotational -- panels (a) and (b) -- and of the circular -- panels (c) and (d) -- swimming gaits of the triangular swimmer. Panels in the left column: trajectories of the triangular swimmer in the $(x, y)$-plane as functions of time (from ``blue'' to ``red'' -- see colour coded time scale next to the panels). For clarity each bead of the swimmer is characterized by a specific colour (as labeled). 
Panels in the right column: visualization of the {\it reciprocal} arm deformations (strokes) for the two different swimming gaits (specified above) over a period of motion; the numbers label the strokes of the periodically repeated deformation of the swimmer (see also text). The size of the beads is not scaled with the arm lengths.
}
\label{fig:triag_trajectories_A}
\end{figure}

\begin{figure}
\centering
\includegraphics[scale = 1.5]{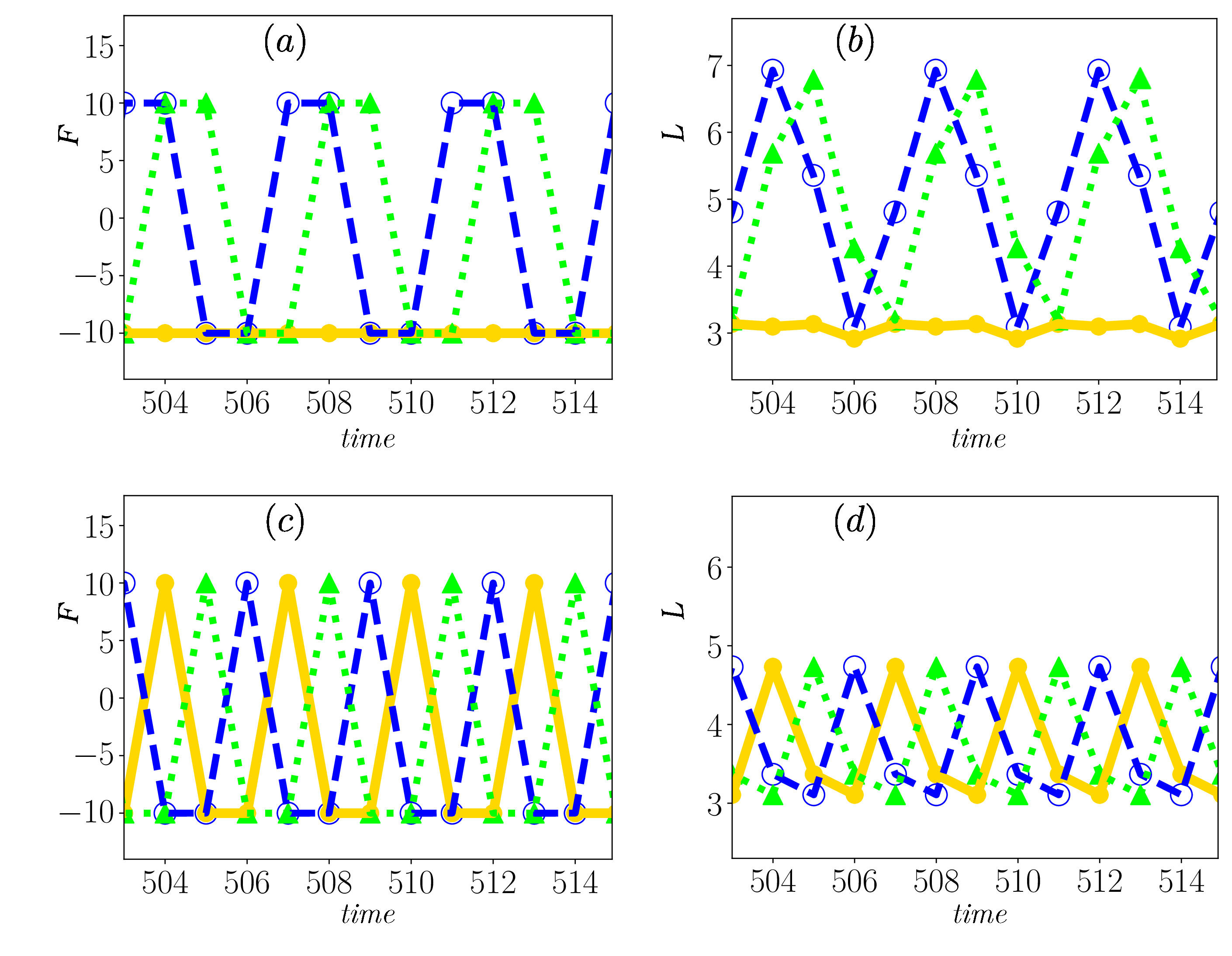}
\caption{Forces $F_{\alpha\beta}$ (panels (a) and (c); as labeled) and arm lengths $L_{\alpha\beta}$ (panels (b) and (d); as labeled) as functions of time over an arbitrarily chosen time window of stationary propagation of the triangular swimmer. Panels in the top row: rotational mode; panels in the bottom row: circular mode. Here, $F_{12}$, $F_{23}$, and $F_{13}$ are indicated by,
%\begin{tikzpicture}[baseline=-0.5ex]
 % \draw[Goldenrod, solid, line width=3pt] (0,0) -- (1cm,0);
%\end{tikzpicture} 
(Yellow, solid),
%\begin{tikzpicture}[baseline=-0.5ex]
 % \draw[blue, dashed, line width=1.9pt] (0,0) -- (1cm,0);
%\end{tikzpicture} 
(blue, dashed), and
%\begin{tikzpicture}[baseline=-0.5ex]
 % \draw[green, dotted, line width=1.5pt] (0,0) -- (1cm,0);
%\end{tikzpicture} 
(green, dotted) lines, respectively.
}
\label{fig:triag_f_l_2d_A}
\end{figure}

\begin{figure}
\centering
\includegraphics[scale = 2.2]{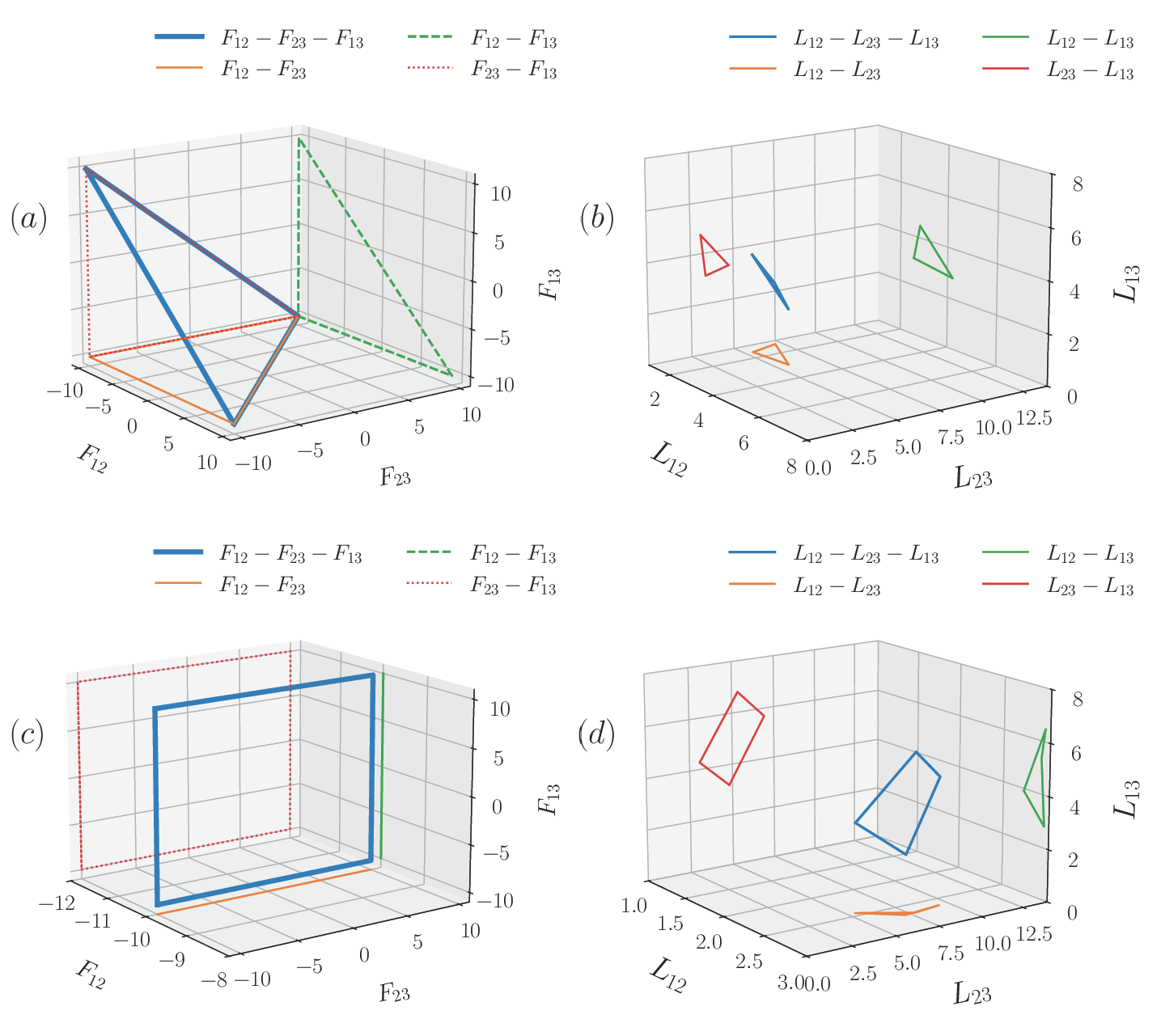}
\caption{Forces $F_{\alpha\beta}$ (panels (a) and (c); as labeled) and arm lengths $L_{\alpha\beta}$ (panels (b) and (d); as labeled) in their respective three-dimensional representations (blue curves). The red, green, and orange curves represent projections of the blue curve onto the related two-dimensiohnal subspaces (as labeled). Curves are shown over a periodically repeated stroke pattern, displayed for an arbitrarily chosen time window of stationary propagation of the triangular swimmer. Panels in the top row: rotational mode, panels in the bottom row: circular mode.
}
\label{fig:triag_f_l_3d_A}
\end{figure}

\begin{figure}
\centering
\includegraphics[scale = 0.9]{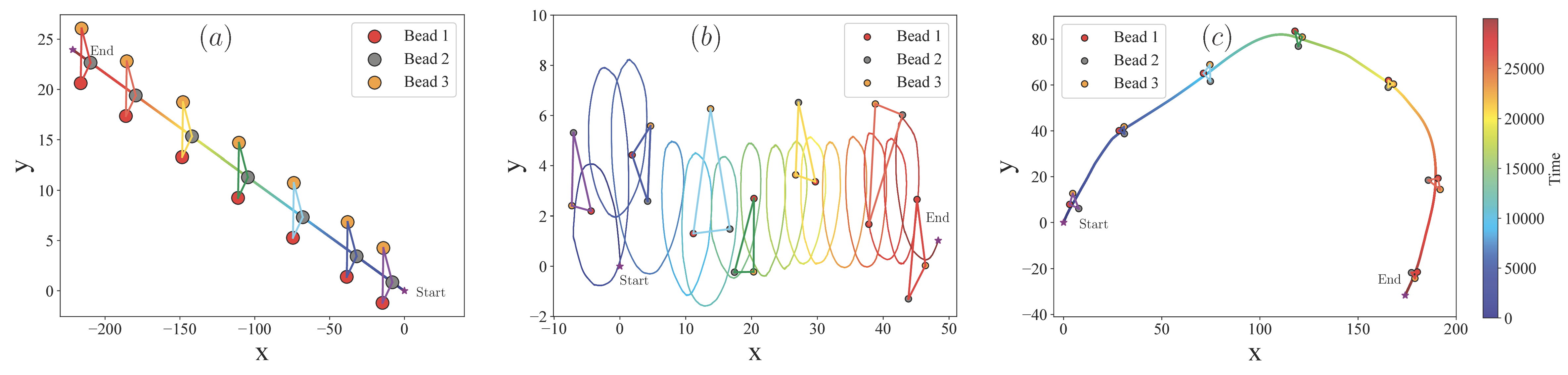}
\caption{Trajectories of the COM of the {\it trained} three-bead swimmer over a time windows of 30~000 time steps (see colour code for the elapsing time $t$ on the right hand side) in the $(x, y)$-plane. Panel (a) -- flapping mode, panel (b) -- chiral mode, panel (c) -- walking mode. Beads are coloured according to the labels. Arm lengths of the swimmer are drawn to the scale while the bead size is not drawn to scale. Positions and configurations of the swimmers are displayed for time instants $t = 1~000, 5~000, 10~000, 15~000, 20~000, 25~000$, and 29~000 time steps. Starting points (at $x = 0$ and $y = 0$) and end points of the trajectories are indicated by symbols.}
\label{fig:triag_long_traj}
\end{figure}

As compared to the rotational mode the situation is somewhat different in the circular mode, where the trajectory of the COM of the swimmer has a circular shape with no net displacement of the swimmer in space. Similar as in the rotational mode the beads orbit around the COM, albeit in a somewhat more complex manner (see panels (c) and (d) of Fig. \ref{fig:triag_trajectories_A}). The mode is built up by four strokes which are repeated periodically. A more detailed analysis of the underlying deformation of the shape of the swimmer along this process can be extracted from the time-dependence of the $F_{\alpha \beta}$ and the $L_{\alpha \beta}$ over a full cycle of strokes: all the three arm lengths ($L_{\alpha \beta}$) show exactly the same time-dependence, with each of the curves being phase-shifted by one stroke. A similar observation can be reported for the $F_{\alpha \beta}$. The fact that all three $L_{\alpha \beta}$ vary in an identical manner as functions of time leads via the aforementioned phase-shift to a circular displacement of the COM. The three-dimensional representations of the $F_{\alpha \beta}$ and the $F_{\alpha \beta}$ show -- along with their two-dimensional projections -- again very simple curves (cf. panels (c) and (d) of Fig. \ref{fig:triag_f_l_3d_A}).

%%%%%%%%%%%%%%%%%%%%%%%%%%%%%%%%%%%%%%%%%%%%%%%%%%%%%%%%%%%%%%%%%%%%%%%%%
%%%%%%%%%%%%%%%%%%%%   CONCLUSIONS (START)      
                        
\section{Conclusions} 
\label{sec:conclusions}

In this contribution we have investigated the training of autonomous swimming strategies of a planar three-bead swimmer under low Reynolds number conditions via a reinforcement learning scheme. This particular model was originally introduced to mimic the features of {\it Chloamydomonas}. In our scheme agent of the swimmer is represented by an artificial neural network (ANN): the three neurons of the input layer are the three arm lengths $L_{\alpha \beta}$ ($\alpha \beta = 1, 2, 3$) while the three neurons of the output layer are the forces acting on the arms $F_{\alpha \beta}$ which -- via the equations-of-motions -- couple again to the arm lengths. The internal architecture of the ANNs (in terms of internal nodes and internodal connections) is optimized. This optimization is based on a flexible and transparent reward scheme which, in turn, is the weighted sum of partial rewards (including translational and rotational movements of the swimmer and their coupling as well as geometric constraints imposed on the shape of the swimmer): in his manner the trained strategies that characterize the different swimming gaits can be identified and interpreted via the structure of the emerging, surprisingly simple ANNs. . The actual training is performed with the help of the ``NeuroEvolution of Augmenting Topologies'' (NEAT) algorithm by tracing the efficiency of the ANNs over NEAT generations. 

By systematically varying the partial reward weights, we identify five different swimming strategies which can be distinguished by the characteristic trajectories of the center-of-mass (COM) of the swimmer. Three of these swimming modes -- termed flapping, chiral, and walking modes -- are non-reciprocal in the shape deformation of the swimmer and thus lead to a net propagation of the COM: depending on the weights of the partial rewards the related displacements are either purely translational or combinations of translational and rotational propagation modes. In contrast, two other swimming modes 
are localized in their displacement: the swimmer either rotates  around its COM (rotational) or the COM performs a circular path (circular). These findings are in complete agreement with the scallop theorem.

A deeper analysis of the $L_{\alpha \beta}$ and the $F_{\alpha \beta}$ as functions of time allows a deeper analysis of the different swimming strategies and offers insight into the decision making strategies of the swimmer. Crucially those agents (or ANNs) that perform best in the spirit of the above mentioned optimization scheme are rather simple in their internal architecture; in some swimming modes the related ANNs are even bare of internal nodes. This feature might be crucial for future applications in micro-scale controllers as the implementation of such simple networks -- especially if reducible to sophisticatedly coupled limit-cycle dynamics -- might be well within reach. In contrast, other, less favorable controllers show a considerably more complex internal architecture of the ANNs. 

Soft microswimmers \cite{yan_colloid} -- capable of morphological deformation -- are widely recognized as promising agents for targeted biomedical applications.
The obtained results are helpful and of practical relevance for designing and training artificial microswimmers to perform specific tasks.
Prior studies, employing Q-learning to train triangular microswimmers \cite{jebellat, hesam}, have generally failed to produce locomotion strategies that are effective for autonomous drug delivery or for replicating biological analogues. In contrast, the present work employs a simplified ANN coupled with a fine-tuned complex reward scheme, enabling the swimmer to adopt distinct and efficient swimming gaits.  
The swimmer developed herein utilizes coordinated shape deformations that enhance motility in complex environments, thereby rendering the system a promising candidate for drug cargo transport applications.

\section{Acknowledgement}
This research was funded in whole or in part by the Austrian Science Fund (FWF) (DOI: 10.55776/ESP382). For open access purposes, the author has applied a CC BY public copyright license to any author-accepted manuscript version arising from this submission. The computational results were obtained by using the Vienna Scientific Cluster (VSC) and the Cloud Infrastructure Platform (CLIP, Vienna Biocenter).

\section{Author Contributions}
R.M. and M.H. have design the study. G.K. has directed the research. R.M. has carried out the calculations and numerical simulations and prepared the figures. All authors have contributed to analyzing the data, interpreting the results, and editing the manuscript.

\section{Ethics declarations}
The authors declare no competing financial interests.

\section{Data Availability}
The data that support the findings of the study are available from the corresponding author upon reasonable request.
%%%%%%%%%%%%%%%%%%%%   CONCLUSIONS (END)                                 
%%%%%%%%%%%%%%%%%%%%%%%%%%%%%%%%%%%%%%%%%%%%%%%%%%%%%%%%%%%%%%%%%%%%%%%%%

\bibliographystyle{unsrt} % We choose the "plain" reference style

%\bibliography{microswimmer}
\bibliographystyle{naturemag}

\end{document}